%% file: letter.tex
\documentclass[showpacs,showkeys,aps,prl,twocolumn,groupedaddress]{revtex4}
\input{macro-gn}

\begin{document}

\title{Far-from-constant mean curvature solutions of Einstein's 
       constraint equations \\ with positive Yamabe metrics}

\author{M.~Holst}
\affiliation{Department of Mathematics,
         University of California San Diego,
         La Jolla CA 92093}

\author{G.~Nagy}
\affiliation{Department of Mathematics,
         University of California San Diego,
         La Jolla CA 92093}

\author{G.~Tsogtgerel}
\affiliation{Department of Mathematics,
         University of California San Diego,
         La Jolla CA 92093}

\date{\today}

\begin{abstract}
   We establish new existence results for the Einstein constraint equations 
   for mean extrinsic curvature arbitrarily far from constant. The results 
   hold for rescaled background metric in the positive Yamabe class, with 
   freely specifiable parts of the data sufficiently small, and with matter 
   energy density not identically zero.  Two technical advances make these 
   results possible: A new topological fixed-point argument without smallness 
   conditions on spatial derivatives of the mean extrinsic curvature, and a 
   new global supersolution construction for the Hamiltonian constraint that 
   is similarly free of such conditions.  The results are presented for 
   strong solutions on closed manifolds, but also hold for weak solutions and
   for compact manifolds with boundary.  These results are apparently the 
   first that do not require smallness conditions on spatial derivatives of 
   the mean extrinsic curvature.
\end{abstract}



\maketitle

\mathbi{Introduction.}
The question of existence of solutions to the Lichnerowicz-York
conformally rescaled Einstein's constraint equations, for an
arbitrarily prescribed mean extrinsic curvature, has remained an open
problem for more than thirty years~\cite{jY73}.  The rescaled
equations, which are a coupled nonlinear elliptic system consisting of
the scalar Hamiltonian constraint coupled to the vector momentum
constraint, have been studied almost exclusively in the setting of
constant mean extrinsic curvature, known as the CMC case. In the CMC
case the equations decouple, and it has long been known how to
establish existence of solutions. The case of CMC data on closed
(compact without boundary) manifolds was completely resolved by
several authors over the last twenty years, with the last remaining
subcases resolved and summarized by Isenberg in~\cite{jI95}. 
Over the last ten years, other CMC cases 
were studied and resolved; see the survey~\cite{rBjI04}.

Conversely, the question of existence of solutions to the Einstein
constraint equations for nonconstant mean extrinsic curvature 
(the ``non-CMC case'') has remained largely unanswered, with progress
made only in the case that the mean extrinsic curvature is nearly
constant (the ``near-CMC case''), in the sense that the
size of its spatial derivatives is sufficiently small.
The near-CMC condition leaves the constraint equations
coupled, but ensures the coupling is weak.  In~\cite{jIvM96}, Isenberg
and Moncrief established the first existence (and uniqueness) result
in the near-CMC case, for background metric having negative Ricci
scalar.  Their result was based on a fixed-point argument, together
with the use of iteration barriers (sub- and supersolutions) 
which were shown to be bounded above and below by fixed positive constants,
independent of the iteration.
We note that
both the fixed-point argument and the global barrier construction
in~\cite{jIvM96} rely critically on the near-CMC assumption.  All
subsequent non-CMC existence results are based on the analysis
framework in~\cite{jIvM96} and are thus limited to the near-CMC case
(see the survey~\cite{rBjI04}, the nonexistence results
in~\cite{jInOM04}, and also the newer existence results
in~\cite{pAaCjI07} for non-negative Yamabe classes).

This article presents the first non-CMC existence results for the
Einstein constraints that do not require the near-CMC assumption.  Two
recent advances make this possible: A new topological
fixed-point argument (established in~\cite{mHjKgN07,mHgNgT07b})
and a new global supersolution construction for the Hamiltonian 
constraint (presented here and in~\cite{mHgNgT07b})
that are both free of near-CMC conditions.
These two results allow us to establish existence of
non-CMC solutions for conformal background metrics in the positive
Yamabe class, with the freely specifiable part of the data given by
the traceless transverse part of the rescaled extrinsic curvature and
the matter fields sufficiently small, and with the matter energy
density not identically zero. We only state the main results and give
the ideas of the proofs; detailed proofs may be found
in~\cite{mHgNgT07b} for closed manifolds and
in~\cite{mHjKgN07} for compact manifolds with boundary.
Our results here and
in~\cite{mHjKgN07,mHgNgT07b} reduce the remaining open questions of
existence of non-CMC solutions without near-CMC conditions to two
basic open questions: (1) Existence of global {\em super}-solutions
for background metrics in the nonpositive Yamabe classes and for
large data; and (2) existence of global {\em sub}-solutions for
background metrics in the positive Yamabe class in vacuum.

\mathbi{The Conformal Method.}
The manifold and fields $(\cM ,\hat h_{ab} ,\hat k^{ab} ,\hat
j^a,\hat\rho)$ form an {\em initial data set} for Einstein's equations
iff $\cM$ is a $3$-dimensional smooth manifold, $\hat h_{ab}$ is a
Riemannian metric on $\cM$, $\hat k^{ab}$ is a symmetric tensor field
on $\cM$, $\hat j^a$ and $\hat\rho$ are a vector field and a
non-negative scalar field on $\cM$, respectively, satisfying an energy
condition (described below), and the following hold on $\cM$,
\begin{align}
\label{hc}
\hat R + \hat k^2 - \hat k_{ab}\hat k^{ab} -2\kappa \hat \rho &=0,\\
\label{mc}
-\hat \nabla_a \hat k^{ab} + \hat\nabla^b\hat k +\kappa \hat j^b &=0.
\end{align}
Here, $\hat\nabla_a$ is the Levi-Civita connection of $\hat h_{ab}$,
so it satisfies $\hat\nabla_a\hat h_{bc} =0$, $\hat R$ is the Ricci
scalar of the connection $\hat\nabla_a$, $\hat k =\hat h_{ab}\hat
k^{ab}$ is the trace of $\hat k^{ab}$, and $\kappa = 8\pi$ in units
where both the gravitational constant and speed of light have value
one.  We denote by $\hat h^{ab}$ the tensor inverse of $\hat
h_{ab}$. Tensor indices of hatted quantities are raised and lowered
with $\hat h^{ab}$ and $\hat h_{ab}$, respectively.
When~(\ref{hc})-(\ref{mc}) hold, the manifold $\cM$ can be embedded as
a hyper-surface in a $4$-dimensional manifold corresponding to a
solution of the space-time Einstein field equations, and the
push-forward of $\hat h^{ab}$ and $\hat k^{ab}$ represent the first
and second fundamental forms of the embedded hyper-surface.  This
leads to the terminology extrinsic curvature for $\hat k^{ab}$, and
mean extrinsic curvature for its trace, $\hat k$.
The dominant energy condition on the matter fields implies
the energy condition $-\hat\rho^2 +\hat h_{ab}\hat j^a\hat j^b \leqs 0$,
with strict inequality at points on $\cM$ where $\rho \neq 0$;
see~\cite{Wald84}. This condition is why the trivial procedure of
fixing an arbitrary Riemannian metric $\hat h_{ab}$ and a symmetric
tensor $\hat k^{ab}$ and then defining $\hat j^a$ and $\hat \rho$
by~(\ref{hc})-(\ref{mc}) does not generally give a physically
meaningful initial data set for Einstein's equations.

The conformal method consists of finding solutions $\hat h_{ab}$,
$\hat k^{ab}$, $\hat j^a$ and $\hat\rho$ of~(\ref{hc})-(\ref{mc})
using a particular decomposition.  To proceed, fix on $\cM$ a
Riemannian metric $h_{ab}$ with Levi-Civita connection $\nabla_a$, so
it satisfies $\nabla_ah_{bc}=0$, and has Ricci scalar $R$. Fix on
$\cM$ a symmetric tensor $\sigma^{ab}$, trace-free and divergence-free
with respect to $h_{ab}$, that is, $h_{ab}\sigma^{ab}=0$ and
$\nabla_a\sigma^{ab}=0$.  Also fix on $\cM$ scalar fields $\tau$ and
$\rho$, and a vector field $j^a$,
subject to the condition $-\rho^2 + h_{ab} j^aj^b \leqs 0$,
with strict inequality at points on $\cM$ where $\rho \neq 0$.
We have denoted by $h^{ab}$ the tensor inverse of $h_{ab}$, and we
use the convention that tensor indices of unhatted quantities are
raised and lowered with the tensors $h^{ab}$ and $h_{ab}$,
respectively. Finally, given a smooth vector field $w^a$ on $\cM$,
introduce the conformal Killing operator $\cL$ as follows, $(\cL
w)^{ab} = \nabla^a w^b + \nabla^b w^a - (2/3)\, (\nabla_cw^c)\,
h^{ab}$. The conformal method then involves first solving the
following equations for a scalar field $\phi$ and vector field $w^a$
\begin{gather}
\label{rhc}
-\Delta\phi + a_{\tiR} \phi +  a_{\tau}\,\phi^{5} 
- a_w \, \phi^{-7} - a_{\rho} \, \phi^{-3}= 0,\\
\label{rmc}
-\nabla_a(\cL w)^{ab} + b_{\tau}^b \phi^{6} + b_j^b=0,
\end{gather}
where we have introduced the Laplace-Beltrami operator
$\Delta\phi=h^{ab}\nabla_a\nabla_b\phi$, and the functions
$a_{\tiR}=R/8$, $a_{\tau}=\tau^2/12$, $a_{\rho}=\kappa \rho/4$,
$b_{\tau}^b=(2/3)\nabla^b \tau$, 
$b_{j}^b=\kappa j^b$, and
$a_{w}= \bigr[\sigma_{ab}+(\cL w)_{ab}\bigr] 
\bigr[\sigma^{ab}+(\cL w)^{ab}\bigr]/8$.
One then recovers the tensors $\hat h_{ab}$, $\hat k^{ab}$, $\hat j^a$
and $\hat\rho$ through the expressions
\begin{gather}
\label{hjr}
\hat h_{ab} = \phi^4 h_{ab},
\quad
\hat j^a = \phi^{-10} \,j^a,
\quad
\hat\rho = \phi^{-8}\,\rho,\\
\label{k}
\hat k^{ab} = \phi^{-10} \bigl[ \sigma^{ab} 
+(\cL w)^{ab}\bigr] + \frac{1}{3} \phi^{-4} \tau \,h^{ab}.
\end{gather}
A straightforward computation shows that if $\hat h_{ab}$, $\hat
k^{ab}$, $\hat j^a$ and $\hat \rho$ have the form given
in~(\ref{hjr})-(\ref{k}), then equations~(\ref{hc})-(\ref{mc}) are
equivalent to~(\ref{rhc})-(\ref{rmc}).
Hatted fields represent quantities with physical meaning,
except the trace $\tau$ of the physical extrinsic curvature $\hat k^{ab}$, 
that is, $\tau= \hat k$.

We employ standard $L^p$ and Sobolev spaces $W^{k,p}$,
following~\cite{Adams75} for scalar-valued functions on bounded sets
in $\R^n$, and following~\cite{Hebey96} and~\cite{Palais68} for
generalizations to manifolds and to tensor fields.  The space
$L^{\infty}$ is the set of almost everywhere (a.e.) bounded functions
on $\cM$, which is a Banach space with norm $\|u\|_{\infty}:=\mbox{ess
}\sup_{\cM}|u|$.  The Banach space $L^p$, with $1\leqs p<\infty$, is
the set of tensor fields on $\cM$ having norm $\|u\|_p:=\bigl[
\int_{\cM}(u_{a_1\cdots a_n} u^{a_1\cdots a_n})^{p/2} dx\bigr]^{1/p}$
finite. The Banach space $W^{k,p}$ is the set of tensor fields on
$\cM$ having $k\geqs 0$ weak covariant derivatives in $L^p$, with norm
denoted $\|~\|_{k,p}$.

\mathbi{The Momentum Constraint.}
The momentum constraint~(\ref{rmc}) is well-understood in the case
that $h_{ab}$ has no conformal Killing vectors (a vector field $v^a$
is {\em conformal Killing} iff $(\cL v)^{ab} =0$).  A standard result
is the following.  Let $(\cM,h_{ab})$ be a 3-dimensional, closed,
$C^2$, Riemannian manifold, with $h_{ab}$ having no conformal Killing
vectors, and let $b_{\tau}^a$, $b_j^a\in L^{p}$ with $p\geqs 2$ and
$\phi\in L^\infty$; Then, equation~(\ref{rmc}) has a unique solution
$w^{a}\in W^{2,p}$ with
\begin{equation}
\label{w-bound}
c\,\|w\|_{2,p} \leqs 
\|\phi\|^6_{\infty} \,\|b_{\tau}\|_{p} + \|b_j\|_{p},
\end{equation}
where $c>0$ is a constant. We have generalized this result
in~\cite{mHjKgN07,mHgNgT07b}, allowing weaker coefficient
differentiability, giving existence of solutions down to $w^a\in
W^{1,p}$, with real number $p\geqs 2$. The proof in~\cite{mHgNgT07b}
is based on Riesz-Schauder theory for compact
operators~\cite{Wloka87}. The case of compact manifold $\cM$ with
boundary is analyzed in~\cite{mHjKgN07}.

From inequality~(\ref{w-bound}) it is not difficult to show that for
$p> 3$ the following pointwise estimate holds,
\begin{equation}
\label{MC-awest}
a_{\biw}\leqs K_1 \, \|\phi\|_{\infty}^{12}+ K_2,
\end{equation}
with $K_1=\frac{1}{2}(\frac{c_s c_{\mathcal{L}}}{c})^2
\|b_{\tau}\|_p^2$, $K_2 =\frac{1}{4}\|\sigma\|_\infty^2 + \frac{1}{2}
(\frac{c_sc_{\mathcal{L}}}{c})^2\|b_j\|_p^2$, where $c_s$ is the
constant in the embedding $W^{1,p}\hookrightarrow L^{\infty}$, and
$c_{\mathcal{L}}$ is a bound on the norm of $\mathcal{L}:W^{2,p}\to
W^{1,p}$. There is no smallness assumption on
$\|b_{\tau}\|_p$, so the near-CMC condition is not required for these
results.

\mathbi{Global Hamiltonian Constraint Barriers.}
Let $\cM$ be closed.  The scalar functions $\phi_{-}$ and $\phi_{+}$
are called {\em barriers} (sub- and supersolutions, respectively) iff
\begin{align}
\label{gsbs}
-\Delta\phi_{-} + a_{\tiR} \phi_{-} +  a_{\tau}\,\phi_{-}^{5} 
- a_w \, \phi_{-}^{-7} - a_{\rho} \, \phi_{-}^{-3} &\leqs 0,\\
\label{gsps}
-\Delta\phi_{+} + a_{\tiR} \phi_{+} +  a_{\tau}\,\phi_{+}^{5} 
- a_w \, \phi_{+}^{-7} - a_{\rho} \, \phi_{+}^{-3} &\geqs 0.
\end{align}
The barriers are {\em compatible} iff $0<\phi_{-}\leqs\phi_{+}$, and
are {\em global} iff~(\ref{gsbs})--(\ref{gsps}) holds for all $w^a$
solving equation~(\ref{rmc}), with source $\phi \in
[\phi_{-},\phi_{+}]$.  The closed interval
\begin{equation}
  \label{eqn:interval}
[\phi_{-},\phi_{+}]=\{\phi \in L^p : \phi_{-} \leqs \phi \leqs \phi_{+} 
\mbox{~a.e. in~} \cM\},
\end{equation}
is a topologically closed subset of $L^p$, $1 \leqs p \leqs \infty$
(see~\cite{mHgNgT07b}).  Global supersolutions are difficult to find
as a consequence of the non-negativity of $a_w$ and its
estimate~(\ref{MC-awest}), together with the
limit~(\ref{eqn:interval}).
All previous global supersolution constructions, such as those
in~\cite{jIvM96,pAaCjI07}, rely in a critical way on the near-CMC
assumption, which appears as the condition that a suitable norm of
$\nabla \tau$ be sufficiently small, or equivalently, that $K_1$
in~(\ref{MC-awest}) be sufficiently small.  The main result in this
letter is to establish existence of global supersolutions of the
Hamiltonian constraint without the near-CMC assumption.  We need the
following notation: Given any scalar function $v\in L^{\infty}$,
denote by $v^{\tiwedge} =\mbox{ess sup}_{\cM}v$, and $v^{\tivee}
=\mbox{ess inf}_{\cM}v$.
\begin{theorem}
\label{T:gsps}
Let $(\cM,h_{ab})$ be a $3$-dimensional, smooth, closed Riemannian
manifold with metric $h_{ab}$ in the positive Yamabe class with no 
conformal Killing vectors.
Let $u$ be a smooth positive solution of the Yamabe problem
\begin{equation}
\label{y}
-\Delta u + a_{\tiR}u - u^5 = 0,
\end{equation}
and define the constant $k= u^{\tiwedge}/u^{\tivee}$.  If the function
$\tau$ is nonconstant and the rescaled matter fields $j^a$, $\rho$,
and traceless transverse tensor $\sigma^{ab}$ are sufficiently small,
then
\begin{equation}
\label{epsilon}
\phi_{+} =\epsilon u,\quad
\epsilon = \Bigl[\frac{1}{2K_1 k^{12}} \Bigr]^{\frac{1}{4}}
\end{equation}
is a global supersolution of equation~(\ref{rhc}).
\end{theorem}

{\em Proof. (Theorem~\ref{T:gsps})} Existence of a smooth positive
solution $u$ to~(\ref{y}) is summarized in~\cite{jLtP87}.  Using the
notation
\begin{equation}
  \label{E:E}
E(\phi)= -\Delta \phi + a_{\tiR} \phi 
+ a_{\tau}\phi^5 - a_w\phi^{-7} - a_{\rho}\phi^{-3},
\end{equation}
we have to show $E(\phi_{+})\geqs 0$.  Taking $\phi_{+}= \epsilon u$,
$\epsilon > 0$ gives the identity $-\Delta \phi_{+} + a_{\tiR}\phi_{+}
= \epsilon u^5$.  We have
\begin{align*}
E(\phi_{+}) &\geqs -\Delta \phi_{+} + a_{\tiR}\phi_{+}
- \frac{K_1 (\phi_{+}^{\tiwedge})^{12} + K_2}{\phi_{+}^{7}}
- \frac{a_{\rho}^{\tiwedge}}{\phi_{+}^{3}}\\
& \geqs \epsilon \,u^5 -  
K_1 \Bigl[\frac{\phi_{+}^{\tiwedge}}{\phi_{+}^{\tivee}}\Bigl]^{12} 
\,\phi_{+}^5 
-\frac{K_2}{\phi_{+}^{7}} - \frac{a_{\rho}^{\tiwedge}}{\phi_{+}^{3}}\\
&\geqs \epsilon u^5 \Bigl[ 
1 - K_1 \, k^{12} \epsilon^4
- \frac{K_2}{\epsilon^8 u^{12}} 
- \frac{a_{\rho}^{\tiwedge}}{\epsilon^{4} u^8} \Bigr],
\end{align*}
where we have used $\phi_{+}^{\tiwedge}/\phi_{+}^{\tivee} =
u^{\tiwedge}/u^{\tivee} = k$.  The choice of $\epsilon$ made
in~(\ref{epsilon}) is equivalent to $1/2 = 1 - K_1 \, k^{12}
\epsilon^4$.  For this $\epsilon$, impose on the free data
$\sigma^{ab}$, $\rho$ and $j^a$ the condition
\[
\frac{1}{2} - \frac{K_2}{\epsilon^8 (u^{\tivee})^{12}} 
- \frac{a_{\rho}^{\tiwedge}}{\epsilon^4 (u^{\tivee})^8} \geqs 0.
\]
Thus for any $K_1>0$, we can guarantee $E(\phi_{+})\geqs 0$ for
sufficiently small $\sigma^{ab}$, $\rho$ and $j^a$.\eqed

Theorem~\ref{T:gsps} shows that global supersolutions $\phi_{+}$ can
be built without using near-CMC conditions by rescaling solutions to
the Yamabe problem~(\ref{y}); the larger~$\|\nabla\tau\|_p$, the
smaller the factor~$\epsilon$.  Existence of the finite positive
constant $k$ appearing in Theorem~\ref{T:gsps} is related to
establishing a Harnack inequality for solutions to the Yamabe problem
(see~\cite{yLlZ04}).  It remains to construct (again, without near-CMC
conditions) a compatible global subsolution satisfying $0 <
\phi_{-}\leqs\phi_{+}$.  We now give a variant of some known
constructions~\cite{nOMjY74,dm05,yCB04}; so
also~\cite{mHjKgN07,mHgNgT07b}.

\begin{theorem}
\label{T:gsus}
Let the assumptions for Theorem~\ref{T:gsps} hold. If also the
rescaled matter energy density $\rho$ is not identically zero, then
there exists a positive global subsolution $\phi_{-}$ of
equation~(\ref{rhc}), compatible with the global supersolution in
Theorem~\ref{T:gsps}, so that it satisfies $0 < \phi_{-} \leqs
\phi_{+}$.
\end{theorem}

{\em Proof. (Theorem~\ref{T:gsus})} Let $a_{\rho}\geqs\zeta>0$ in some
open set $B\subset\cM$.  We know from \cite{jI95} that there exists
$u$ satisfying
\begin{equation}
\textstyle
\label{yy}
-\Delta u + a_{\tiR}u - R_u u^5 = 0,
\end{equation}
such that $R_u\leqs-\xi<0$ in $\cM\setminus B$.  Taking $\phi_{-}=
\epsilon u$, $\epsilon > 0$ gives the identity $-\Delta \phi_{-} +
a_{\tiR}\phi_{-} = \epsilon R_u u^5$.  Using $E(\phi)$
from~(\ref{E:E}), we must show $E(\phi_{-})\leqs 0$.  We have
\begin{align*}
E(\phi_{-}) & = -\Delta \phi_{-} + a_{\tiR} \phi_{-} 
+ a_{\tau}\phi_{-}^5 - a_w\phi_{-}^{-7} - a_{\rho}\phi_{-}^{-3}\\
& \leqs \epsilon R_{u} (u^{\tivee})^5
+ a_{\tau}^{\tiwedge} \epsilon^5 (u^{\tiwedge})^5 
- a_{\rho} \epsilon^{-3} (u^{\tivee})^{-3}.
\end{align*}
Now find $\epsilon=\epsilon_1>0$ sufficiently small so on
$B\subset\cM$,
\[
\epsilon_1 R_{u} (u^{\tivee})^5
+ a_{\tau}^{\tiwedge} \epsilon_1^5 (u^{\tiwedge})^5 
- \zeta \epsilon_1^{-3} (u^{\tivee})^{-3} \leqs 0.
\]
Next find $\epsilon=\epsilon_2>0$ sufficiently small so on
$\cM\setminus B$,
\[
- \xi \epsilon_2 (u^{\tivee})^5
+ a_{\tau}^{\tiwedge} \epsilon_2^5 (u^{\tiwedge})^5 \leqs 0.
\]
Taking now $\epsilon_0=\min\{\epsilon_1,\epsilon_2\}>0$ produces a
global subsolution $\phi_{-} = \epsilon u$, for any $\epsilon \in
(0,\epsilon_0]$.  We now finally take $\epsilon \in (0,\epsilon_0]$
sufficiently small so that $0 < \phi_{-} \leqs \phi_{+}$.
\eqed

\mathbi{The Hamiltonian Constraint.}
We now state some supporting results we need
from~\cite{mHjKgN07,mHgNgT07b} for solutions of~(\ref{rhc}).  We state
only the results for strong solutions, recovering previous results
in~\cite{jIvM96,jI95}. Generalizations allowing weaker
differentiability conditions on the coefficients appear
in~\cite{mHjKgN07,mHgNgT07b}.
\begin{theorem}
\label{T:HC}
Let $(\cM,h_{ab})$ be a $3$-dimensional, $C^2$, closed Riemannian
manifold. Let the free data $\tau^2$, $\sigma^2$ and $\rho$ be in
$L^p$, with $p\geqs 2$. Let $\phi_{-}$ and $\phi_{+}$ be barriers
to~(\ref{rhc}) for a particular value of the vector $w^a\in
W^{1,2p}$. Then, there exists a solution $\phi\in
[\phi_{-},\phi_{+}]\cap W^{2,p}$ of the Hamiltonian
constraint~(\ref{rhc}).  Furthermore, if the metric $h_{ab}$ is in the
non-negative Yamabe classes, then $\phi$ is unique.
\end{theorem}
{\em Proof. (Theorem~\ref{T:HC})} The proofs
in~\cite{mHjKgN07,mHgNgT07b} make use of barriers, a priori estimates,
and variational methods.
\eqed

\mathbi{The Coupled Constraint System.}
Our main result concerning the coupled constraint system is the
following.
\begin{theorem}
\label{T:f-p}
Let $(\cM,h_{ab})$ be a $3$-dimensional, smooth, closed Riemannian
manifold with metric $h_{ab}$ in the positive Yamabe class with no
conformal Killing vectors. Let $p>3$ and let $\tau$ be in $W^{1,p}$.
Let $\sigma^2$, $j^a$, and $\rho$ be in $L^p$ and satisfy the
assumptions for Theorems~\ref{T:gsps} and~\ref{T:gsus} to yield a
compatible pair of global barriers $0 < \phi_{-} \leqs \phi_{+}$ to
the Hamiltonian constraint~(\ref{rhc}).  Then, there exists a scalar
field $\phi\in[\phi_-,\phi_+]\cap W^{2,p}$ and a vector field $w^a\in
W^{2,p}$ solving the constraint equations~(\ref{rhc})-(\ref{rmc}).
\end{theorem}

Theorem~\ref{T:f-p} can be proven using the following topological
fixed-point result established in~\cite{mHjKgN07,mHgNgT07b}.  For a
review of reflexive and ordered Banach spaces,
see~\cite{mHjKgN07,mHgNgT07b,Zeidler-I}.  Note that such compactness
arguments do not give uniqueness.

\begin{lemma}
\label{L:gf-p}
Let $X$ and $Y$ be Banach spaces, and let $Z$ be a Banach
space with compact embedding $X \hookrightarrow Z$.  Let $U \subset Z$
be a nonempty, convex, bounded subset which is closed in the topology of $Z$,
and let the maps
\[
S: U \to \mathcal{R}(S) \subset Y,
\quad
\quad
T:U \times \mathcal{R}(S) \to U \cap X,
\]
be continuous.  
Then there exist $w\in\mathcal{R}(S)$ and $\phi\in U \cap X$ such that
\begin{equation}
\label{E:fp}
\phi=T(\phi,w)\quad\textrm{and}\quad w=S(\phi).
\end{equation}
\end{lemma}
{\em Proof. (Lemma~\ref{L:gf-p})} The proofs of this result and
several useful variations appear in~\cite{mHjKgN07,mHgNgT07b}.
\eqed

{\em Proof. (Theorem~\ref{T:f-p})} The proof is through
Lemma~\ref{L:gf-p}.  First, for arbitrary real number $s>0$, express
(\ref{rhc})-(\ref{rmc}) as
\begin{equation}
\label{sc}
L_s\phi+f_s(\phi,w)=0,\quad
(\biL w)^a + \bif(\phi)^a =0,
\end{equation}
where $L_s: W^{2,p}\to L^p$ and $\biL :W^{2,p}\to L^p$ are defined as
$L_s\phi := [-\Delta + s] \phi$, and $(\biL w)^a := -\nabla_b(\cL
w)^{ab}$, and where $f_s:[\phi_{-},\phi_{+}]\times W^{2,p} \to L^p$
and $\bif:[\phi_{-},\phi_{+}]\to L^p$ are
\begin{align*}
f_s(\phi,w) &:= [a_{\tiR}-s]\phi + a_{\tau}\phi^5-a_w\phi^{-7} 
- a_{\rho}\phi^{-3},\\
\bif(\phi)^a &:= b_{\tau}^a \phi^6 + b_{j}^a.
\end{align*}
Introduce now the operators $S:[\phi_{-},\phi_{+}]\to W^{2,p}$ and
$T:[\phi_{-},\phi_{+}]\times W^{2,p}\to W^{2,p}$ which are given by
\[
S(\phi)^a := -[\biL^{-1}\bif(\phi)]^a,\quad
T(\phi,w) := -L_s^{-1}f_s(\phi,w).
\]
The mappings $S$ and $T$ are well-defined due to the absence of conformal
Killing vectors and by introduction of the positive shift $s>0$, 
ensuring both $\biL$ and $L_s$ are invertible
(see~\cite{mHgNgT07b,mHjKgN07}).
The 
equations~(\ref{sc}) have the form~(\ref{E:fp}) for use of
Lemma~\ref{L:gf-p}.  We have the Banach spaces $X=W^{2,p}$
and $Y=W^{2,p}$, and the (ordered) Banach space $Z=L^{\infty}$ with
compact embedding $W^{2,p}\hookrightarrow L^{\infty}$.  
The compatible barriers form the nonempty, convex, bounded
$L^{\infty}$-interval $U=[\phi_-,\phi_+]$,
which we noted earlier is closed in $L^p$ for $1 \leqs p \leqs \infty$
(see~\cite{mHgNgT07b}).
It remains to show $S$ and $T$
are continuous maps.
These properties follow from
equation~(\ref{w-bound}) and from Theorem~\ref{T:HC} with global
barriers from Theorem~\ref{T:gsps} and Theorem~\ref{T:gsus},
using standard inequalities.
Theorem~\ref{T:f-p} now follows from Lemma~\ref{L:gf-p}.
\eqed
\\
See~\cite{mHgNgT07b} for generalizations of Theorem~\ref{T:f-p} to
arbitrary space dimensions and allowing weaker differentiability
conditions on the coefficients, establishing existence of nonvacuum,
non-CMC weak solutions down to $\phi\in W^{s,p}$,
for $(s-1)p > 3$.  Generalizations of the
results here and in~\cite{mHgNgT07b} to compact manifolds with
boundary appear in~\cite{mHjKgN07}.

\begin{acknowledgments}
MH was supported in part by NSF Grants~0715145, 0411723, and 0511766,
and DOE Grants DE-FG02-05ER25707 and DE-FG02-04ER25620.  GN and GT
were supported in part by NSF Grants~0715145 and 0411723.
\end{acknowledgments}

\bibliography{ref-gn}

\end{document}

%% file: macro-gn.tex
\usepackage{array,multirow}
\usepackage{amsmath}
\usepackage{amssymb}
\usepackage{amsfonts}
\usepackage{graphicx}
\usepackage{bm}
\usepackage{enumerate}

\newtheorem{theorem}{Theorem}
\newtheorem{lemma}{Lemma}

\newcounter{mnote}
\newcommand{\eqed}{\hfill $\fbox{\hspace{0.3mm}}$ \vspace{.3cm}}

\newcommand{\leqs}{\leqslant}      
     
\newcommand{\geqs}{\geqslant}      
      
\newcommand{\tiR}{\mbox{{\tiny $R$}}}

\newcommand{\tiwedge}{\mbox{{\tiny $\wedge$}}}
\newcommand{\tivee}{\mbox{{\tiny $\vee$}}}

\newcommand{\R}{{\mathbb R}}       

\newcommand{\cL}{{\mathcal L}}
\newcommand{\cM}{{\mathcal M}}

\def\mathbi#1{\textbf{\em #1}}

\newcommand{\biL}{\mathbi{L}}

\newcommand{\bif}{\mathbi{f\,}}

\newcommand{\biw}{\mathbi{w}}

